\documentclass[12pt,oneside]{article}
\usepackage[left=3.0cm,top=2cm,bottom=2.5cm,right=3.0cm,head=0cm,foot=0.7cm]{geometry}
\usepackage{graphics,amsmath,amssymb}
\newcommand{\field}[1]{\mathbb{#1}}
\begin{document}
\title{\Large{What Are the New Implications of Chaos for Unpredictability?}}
\author{Charlotte Werndl\\ \normalsize{The Queen's College, University of Oxford}}
\date{\normalsize{This is a pre-copyedited, author-produced PDF of an article accepted for publication in The British Journal for the Philosophy of Science following peer review. The definitive publisher-authenticated version ``C. Werndl (2009), What Are the New Implications of Chaos for Unpredictability, The British Journal for the Philosophy of Science 60 195--220'' is available online at: http://bjps.oxfordjournals.org/content/60/1/195.abstract.}}
\maketitle
\begin{abstract}

From the beginning of chaos research until today, the unpredictability of chaos has been a central theme. It is widely believed and claimed by philosophers, mathematicians and physicists alike that chaos has a new implication for unpredictability, meaning that chaotic systems are unpredictable in a way that other deterministic systems are not. Hence one might expect that the question `What are the new implications of chaos for unpredictability?' has already been answered in a satisfactory way. However, this is not the case. I will critically evaluate the existing answers and argue that they do not fit the bill.
Then I will approach this question by showing that chaos can be defined via mixing, which has never before been explicitly argued for. Based on this insight, I will propose that the sought-after new implication of chaos for unpredictability is the following: for predicting any event all sufficiently past events are approximately probabilistically irrelevant.

\end{abstract}

\hyphenation{predictabi-lity in-tuitive Er-kenntnis Philo-sophy Un-pre-dicta-bility macro-predict-ability micro-unpredict-ability}
\newpage


\tableofcontents

\newpage
\section{Introduction}

In the past decades much ado has been made about chaos research, which has been hailed as having led to revolutionary scientific insights. Since the beginnings of systematically investigating chaos until today, the unpredictability of chaotic systems has been at the centre of interest.

There is widespread belief in the philosophy, mathematics and physics communities (and it has been claimed in various articles and books) that \textit{there is a new implication of chaos for unpredictability, meaning that chaotic systems are unpredictable in a way other deterministic systems are not}. More specifically, what is usually believed is that there is \textit{at least one} new implication of chaos for unpredictability that holds true in \textit{all} chaotic systems.

The physicist James Lighthill, commenting on the impact of chaos on unpredictability, expresses this point as follows:
\begin{quote}
\small{We are all deeply conscious today that the enthusiasm of our forebears for the marvellous achievements of Newtonian mechanics led them to make generalizations in this area of predictability which, indeed, we may have generally tended to believe before 1960, but which we now recognize were false
(Lighthill [1986], p.~38). \\These features connected with predictability that I shall describe from now on, then, are characteristic of absolutely all chaotic systems (\textit{Ibid.}, p.~42).}
\end{quote}
Similarly, Weingartner ([1996], p.~50) says that `the new discovery now was that [...] a dynamical system obeying Newton's laws [...] can become chaotic in its behaviour and practically unpredictable'.

Thus the question \textit{`What are the new implications of chaos for unpredictability?'}\ appears natural, and one might well suppose that it has already been satisfactorily answered. However, this is not the case. On the contrary, there is a lot of confusion about what exactly the new implications of chaos for unpredictability are. Several answers have been proposed, but, as we will see, none of them fit the bill.

Fundamental questions about the limits of predictability have always been of concern to philosophy. So the widespread belief and the various flawed accounts about the new implications of chaos for unpredictability demand clarification. The aim of this paper is to critically discuss existing accounts and to propose a novel and more satisfactory answer.

My answer will be based on two insights. First, I will show that chaos can be defined in terms of mixing. Although mixing is occasionally mentioned in connection with chaos, to the best of my knowledge, so far no one has explicitly argued that chaos can be thus defined. Second, I will argue that mixing has a natural interpretation as a particular form of approximate probabilistic irrelevance which is a form of unpredictability.
On this basis I will propose a general novel answer:  a new implication of chaos for unpredictability is that for predicting any event at any level of precision, all sufficiently past events are approximately probabilistically irrelevant.

The structure of the paper is as follows. Section 2 will provide the background of our discussion. I will introduce dynamical systems, and I will discuss the concepts of unpredictability relevant for this paper.
Section 3 will be about chaos. Here I will show that chaos can be defined in terms of mixing. After that, in section 4 I will examine the existing answers to
the question of the new implications of chaos for unpredictability, which I dismiss as mistaken. In section 5 I propose a general answer that does not suffer from the shortcomings of the other accounts.

\section{Dynamical Systems and Unpredictability}
\subsection{Dynamical Systems}\label{DS}
Chaos is discussed in dynamical systems theory. A dynamical system is a mathematical model consisting of a \textit{phase space} $X$, the set of all possible states of the system, and \textit{evolution equations} that describe how solutions evolve in phase space. Dynamical systems often model natural systems (e.g.\ in the sciences).

There are \textit{discrete dynamical systems} and \textit{continuous dynamical systems}. Discrete dynamical systems are systems in which the time increases in discrete steps. Formally, they consist of a set $X$ as phase space and a map $T:X\rightarrow X$ as evolution equation; the dynamics of the system is given by $x_{n+1}=T(x_{n}),\,\, x_{0} \in X,\,\,n \in \field{N}_{0}$. The solution through $x$ is the sequence $(T^{n}(x))_{n\geq 0}$, which is also referred to as the iterates of $x$. If $T$ is invertible (noninvertible), I speak of an invertible (noninvertible) discrete dynamical system, respectively.
Continuous dynamical systems involve a continuous time parameter. They typically arise from differential equations. By definition, all dynamical systems and thus chaotic systems are deterministic.\footnote{According to the conventional definition of Montague ([1962]) and Earman ([1971]), a dynamical system is deterministic if and only if any two solutions that agree at one time agree at all future times.}

For simplicity I will often confine my attention to discrete dynamical systems. I can do this without loss of generality because all definitions of chaos I will be using can be directly carried over to continuous dynamical systems. Alternatively, a continuous dynamical system can be regarded as chaotic if and only if there is a suitable Poincar\'{e} section such that the discrete dynamical system defined by the Poincar\'{e} map is chaotic (e.g.\ Smith [1998], pp.~92--3).
Hence everything I will say about the new implications of chaos for unpredictability equally applies to continuous dynamical systems.

Dynamical systems divide into two groups: \textit{volume-preserving} systems, among them Hamiltonian systems, and \textit{dissipative} systems.
A volume-preser-ving system is defined as a system in which the phase-space volume is preserved under time evolution, i.e.\ the volume (formally the Lebesgue measure) of any region of phase space remains the same as this region is evolved according to the evolution equations (Smith [1998], p.~16). Dissipative systems are systems which are not volume-preserving.

There are two types of dynamical systems relevant for our discussion. First, if for a discrete system there is a metric
$d$, where $d$ measures the distance between points in phase space, $(X,d,T)$ is called a \textit{`topological dynamical system'}.  It is generally assumed in the literature (e.g.\ Devaney [1986], p.~51), that \textit{topological systems provide a possible framework for characterising chaos}. This makes intuitive sense because it is often imagined that in case of chaotic behaviour there is some way of measuring the distance between points in the phase space $X$ and thus that there is a metric defined on $X$. Moreover, to the best of my knowledge, there is always a natural metric for paradigmatic chaotic systems. Often the phase space is simply a subset of $\field{R}^{n}$, $n\geq 1$, and the metric is the standard Euclidean metric.

The second type of dynamical system is a measure-theoretic dynamical system. It is a quadruple $(X, \Sigma, \mu, T)$ consisting of a phase space $X$, a \linebreak[3] $\sigma$-algebra $\Sigma$ on $X$, a measure $\mu$ with $\mu(X)=1$ and a surjective measurable map $T: X\rightarrow X$. If a property holds for all points in a subset $\bar{X}$ of $X$ for which $\mu(\bar{X})=1$, it is said that it holds for \textit{almost all} points.

Important for us is what is called a \textit{`measure-preserving dynamical system'}. It is a measure-theoretic system where for all $A \in \Sigma$
\begin{equation}\label{invariant}
\mu(T^{-1}(A))=\mu(A),
\end{equation}
where $T^{-1}(A)=\{x\in X:T(x) \in A\}$ (cf.~Cornfeld et al.\ [1982], pp.~3--5). Condition (\ref{invariant}) says that the measure $\mu$ is \textit{invariant} under the dynamics of the system.
Although there exist evolution equations that do not have invariant measures, for very wide classes of systems invariant measures can be proven to exist. For instance, if $T$ is a continuous map on a compact phase space endowed with a metric, there exists at least one invariant measure (Ma\~{n}\'{e} [1983], p.~52).\footnote{Descriptions of a dynamical system via metric spaces and measures are usually related in the following way:
the $\sigma$-algebra $\Sigma$ of a measure-theoretic system is, or at least includes, the Borel $\sigma$-algebra of the metric space $(X,d)$ of the topological  system. The Borel $\sigma$-algebra of $(X,d)$ is the $\sigma$-algebra generated by all open sets of $X$ (cf.~Ma\~{n}\'{e} [1983], pp.~2--3). Intuitively, it is the $\sigma$-algebra which arises from the metric space $(X,d)$.}

As it is sometimes claimed (e.g.\ Eckmann and Ruelle [1985]), \textit{measure-preserving systems provide a possible framework for characterising chaos}. For volume-preserving systems the natural invariant measure is typically the Lebesgue measure or a normalized Lebesgue measure, e.g.\ the microcanonical measure of classical statistical mechanics. For dissipative systems, to the best of my knowledge, all systems that have ever been identified as chaotic have or are supposed to have a natural invariant measure if one considers the following.

Many chaotic systems have attractors. For a topological system $(Y,d,T)$ the set $\Lambda\subset Y$ is an \textit{attractor} if and only if (i) $T(\Lambda)=\Lambda$; \linebreak[3] (ii) there is a neighbourhood $U \supset \Lambda$ such that all solutions are attracted by $\Lambda$, i.e.\ for all $y$ in $U$ $\lim_{n\rightarrow \infty}\inf\{d(T^{n}(y),x)\,|\,x\in \Lambda\}=0$; and (iii) no proper subset of $\Lambda$ satisfies (i) and (ii).
Liouville's theorem implies that only dissipative systems can have attractors (Schuster and Just [2005], p.~\nolinebreak 162).\footnote{Some other definitions of `attractor' allow that volume-preserving systems can have attractors; yet these definitions are not standard in our context.} As we will see in the next section, for chaotic systems the evolution of any bundle of initial conditions eventually enters every region in phase space. This is impossible for the motion approaching an attractor since the attracted solutions never return to where they originated. Hence chaotic behaviour can only occur on $\Lambda$. The chaotic motion is described by a system with phase space $\Lambda$, and the invariant measure is only defined on $\Lambda$. Generally, an attractor on which the motion is chaotic is called a `\textit{strange attractor}'.

Of course, in practice one is often concerned with solutions approaching a strange attractor. Yet after a sufficiently long duration either the solutions enter the attractor or come arbitrarily near to the attractor. In the latter case since the dynamics is typically continuous, when the solutions are sufficiently near to the attractor, they essentially behave like the solutions on the attractor. And in applications such solutions which are sufficiently near to a strange attractor are considered to be chaotic for practical purposes. In particular, in the latter case the unpredictability of solutions very near to the attractor is practically indistinguishable from the one on the attractor. Consequently, for characterising the unpredictability of motion dominated by strange attractors, it is widely acknowledged that \textit{it suffices to consider the dynamics on attractors}, where natural invariant measures can be defined.

\subsection{Natural Invariant Measures}

What are \textit{natural} invariant measures, in particular for dissipative systems? From an observational viewpoint it is natural to demand that the long-run time-averages of almost all solutions approximate the measure. Such measures are called \textit{`physical measures'}. Let us look at them in more detail (cf.~Eckmann and Ruelle [1985], p.~626 and pp.~639--40).

For measure-preserving systems $(X,\Sigma,\mu,T)$ with $\lambda(X)>0$, where $\lambda$ is the Lebesgue measure, the following method identifies physical measures. (M1) (i) Take any $A\subseteq X$. (ii) Take an initial condition $x\in X$. (iii) Consider $L_{A}(x)$, the long-run average of the fraction of iterates of $x$ which are in $A$. (iv) Consider $G_{A}=\{x\in X\,|\,L_{A}(x)=\mu(A)\}$. Then $\mu$ is a physical measure if and only if for any $A\in \Sigma$ Lebesgue-almost all initial conditions approximate the measure of $A$, i.e.\ $\lambda(G_{A})=\lambda(X)$. If such a measure exists, it is unique.

What are physical measures for strange attractors? I will be concerned with two kinds of strange attractors: first, the case where all solutions eventually enter an attractor $\Lambda$ with $\lambda(\Lambda)>0$. Clearly, here method (M1) can be applied for $X=\Lambda$. Second, it can be that the solutions approach but never enter an attractor $\Lambda$ with $\lambda(\Lambda)=0$ but $\lambda(U)>0$, where $U$ is the neighbourhood of $\Lambda$. Here the method has to be slightly modified. (M2): (i) Take any region $A\subseteq \Lambda$.
(ii) Take an initial condition $x\in U$.
(iii) Consider $\bar{L}_{A}(x)$, the long-run average of the fraction of iterates of $x$ which are \textit{close to} $A$.
(iv) Consider $\bar{G}_{A}=\{x\in U\,|\,\bar{L}_{A}(x)=\mu(A)\}$. Then $\mu$ is a physical measure if and only if for all $A\in\Sigma$ it holds that $\lambda(\bar{G}_{A})=\lambda(U)$. If such a measure exists, it is unique.

As we will see in the next section, chaotic systems are ergodic. A measure-preserving system $(X,\Sigma,\mu,T)$ is \textit{ergodic} if and only if for all $A\in \Sigma$ with $\mu(A)>0$:
\begin{equation}\label{ergodic}
\mu(\cup_{n\geq 0}T^{-n}(A))=1.
\end{equation}
Now for ergodic volume-preserving systems the Lebesgue-measure is the physical measure. As we will see in the next section, typically for systems proven to be chaotic physical measures can be proven to exist (Lyubich [2002]; Young [2002]). For system only conjectured to be chaotic numerical evidence generally favours the existence of physical measures (Young [1997]).

\begin{figure}\centering \includegraphics{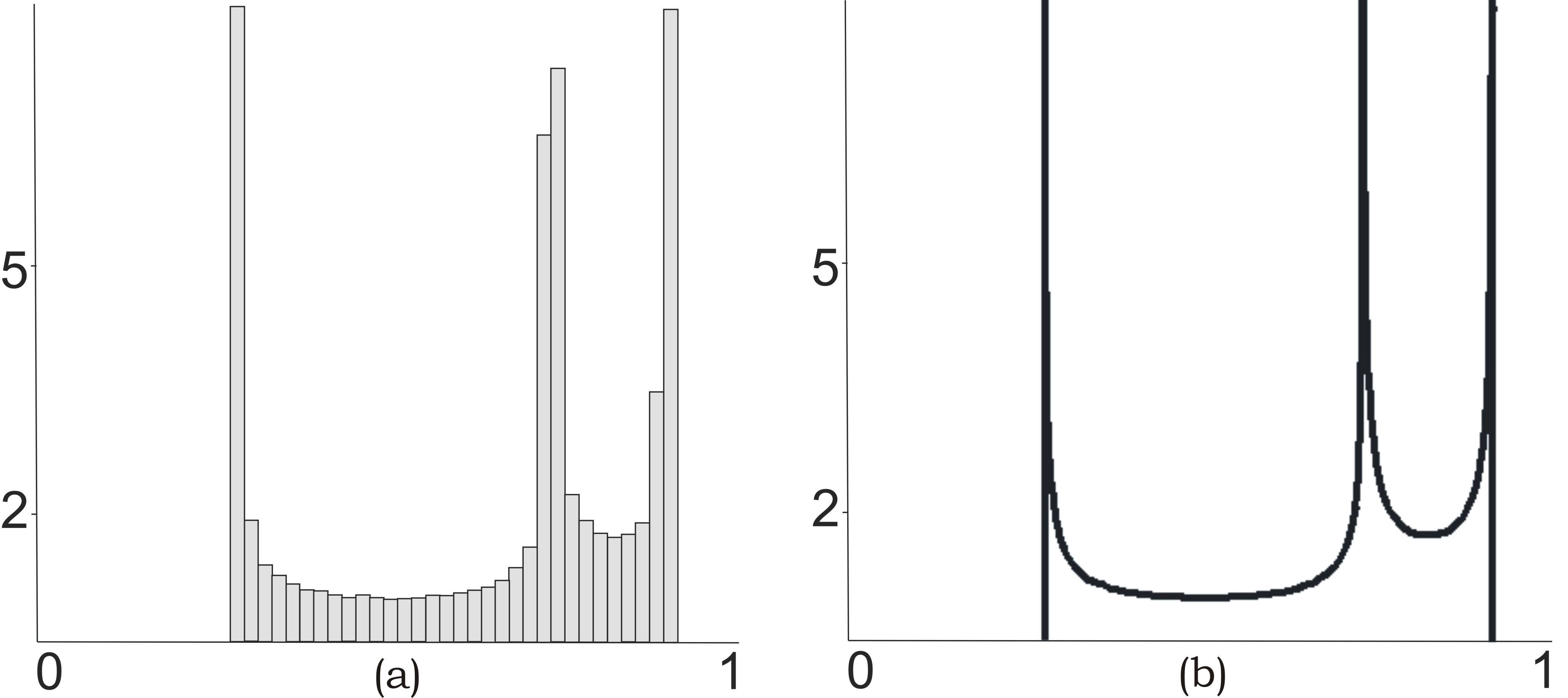} \caption{\small{(a)$\,$histogram and (b)$\,$natural measure of the logistic map for $\alpha\approx 3.6785$}\label{PM}} \end{figure}
For an example consider the logistic map $T(x):[0,1]\rightarrow [0,1]$, $T(x)=\alpha x (1-x)$ with $\alpha\approx 3.6785$. Here the solutions enter an attractor of positive Lebesgue measure. Now we choose an initial condition on the attractor and draw a histogram of the fraction of iterates of $x$ (up to an iterate $T^{n}(x), n\geq 1$) which are in a particular part on the attractor. Then, for Lebesgue-almost all initial conditions we chose on the attractor, we obtain what is illustrated in Figure \ref{PM}: as $n$ goes to infinity and the histogram becomes finer, the histograms approximate a particular measure on the attractor. Hence this measure is physical according to method (M1) (cf.~Jacobson [1981]).

\begin{figure}
\centering
\includegraphics{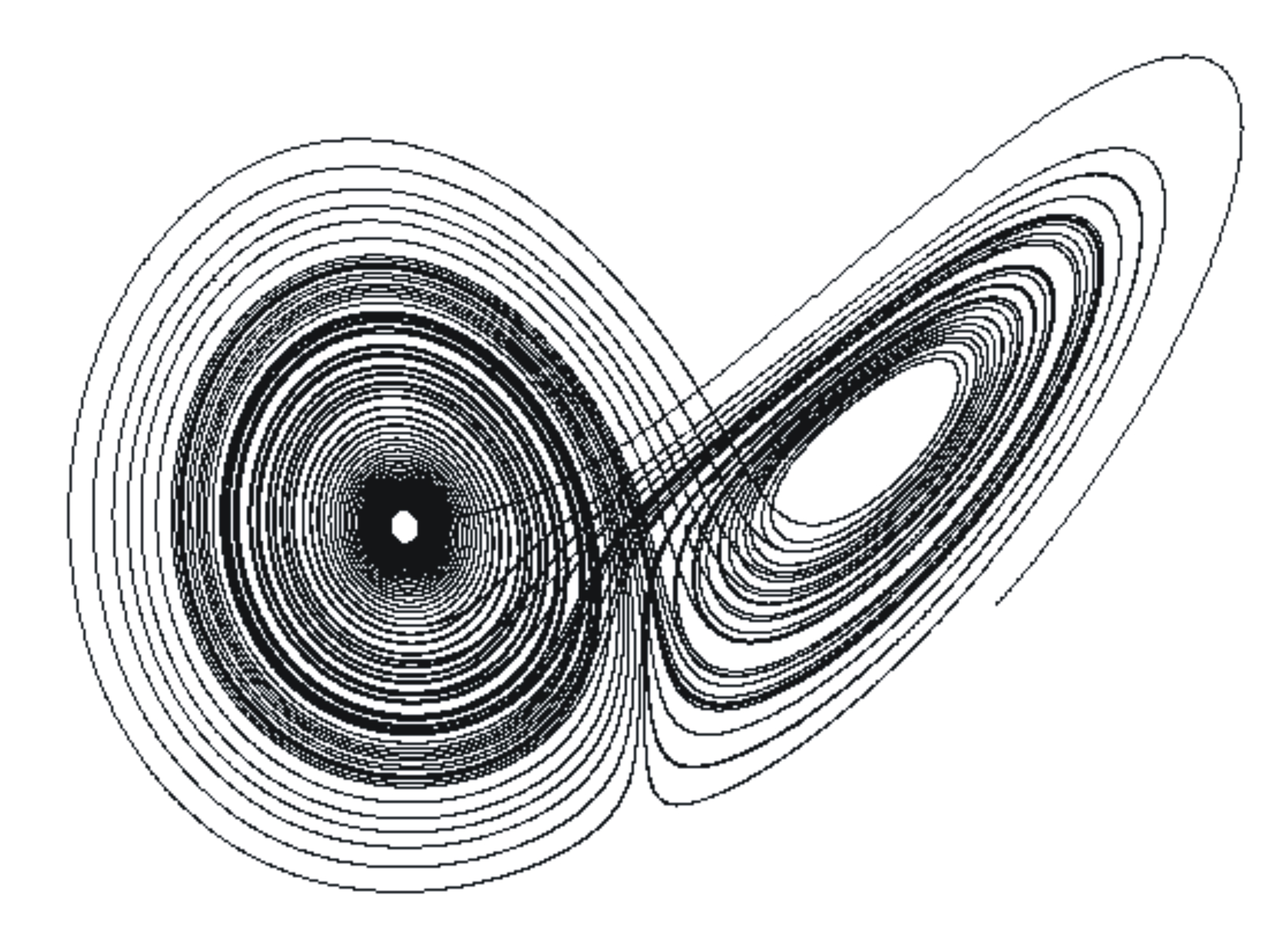}
\caption{\small{Numerical solution of the Lorenz equations for $\sigma = 10$, $r=28$, $b=8/3$}\label{LA}}
\end{figure}

For another example consider the Lorenz equations
\begin{eqnarray}\label{LEO}
\frac{dx(t)}{dt}&=&\sigma(y(t)-x(t))\nonumber\\
\frac{dy(t)}{dt}&=&r x(t)-y(t)-x(t)z(t)\\
\frac{dz(t)}{dt}&=&x(t)y(t)-bz(t)\nonumber,
\end{eqnarray}
for the parameter values $\sigma = 10$, $r=28$ and $b=8/3$, which Lorenz ([1963]) considered. Here it is proven that there is a strange attractor of Lebesgue measure zero such that all solutions originating in the neighbourhood of the attractor, which is of positive Lebesgue measure, approach but never enter the system.
Figure \ref{LA} shows a numerical solution of these equations; one can vaguely discern the shape of the attractor, known as the Lorenz attractor, because the solution spirals toward it.
According to the method (M2), the physical measure is the one for which for the following condition holds: for Lebesgue-almost-all initial conditions in the neighbourhood of the attractor
the long-run time-average a solution is \textit{close} to a set $A$ on the attractor approximates the measure of $A$ (cf.~Luzzatto et al.\ [2005]).\footnote{
There are also other natural measures. For instance, $\nu$ is \textit{absolutely continuous} with respect to $\mu$, where $\nu$ and $\mu$ are measures on a measurable space $(X,\Sigma)$, if and only if for all $A\in\Sigma$ with $\mu(A)=0$ also $\nu(A)=0$. Absolute continuity with respect to the Lebesgue measure can be justified (Malament and Zabell [1980]; van Lith [2001], p.~590). Hence if there is a unique ergodic invariant measure absolutely continuous with respect to the Lebesgue measure, it is a natural one. For ergodic volume-preserving systems the Lebesgue measure is such a unique measure. For many systems, e.g.\ wide classes of one-dimensional maps and, as we will see in the next section, many paradigmatic dissipative chaotic systems including strange attractors, there is a unique ergodic measure absolutely continuous with respect to the Lebesgue measure (Lyubich [2002]). For instance, for the logistic map with $\mu\approx 3.6785$ the measure of Figure \ref{PM}(b) is such a unique measure (Jacobson [1981]).}

Invariant measures are commonly interpreted as probability densities. This deep and controversial issue has, of course, been discussed in statistical mechanics but  is not the main focus of this paper. I only mention two interpretations that naturally suggest interpreting measures as probability and relate to our discussion. According to the \textit{time-average interpretation}, the measure of a set $A$ is the long-run time-average a solution spends in $A$. According to the \textit{ensemble interpretation}, the measure of a set $A$ at $t$ corresponds to the fraction of solutions starting from some set of initial conditions that are in $A$ at time $t$ (Berkovitz et al.\ [2006], p.\ 675).

\subsection{Unpredictability}\label{unpredictability}
There are different kinds of unpredictability in dynamical systems. I will only introduce two concepts needed for the discussion of our main question.

According to the \textit{first} concept of unpredictability, a system is unpredictable when \textit{any bundle of initial conditions spreads out more than a specific diameter representing the prediction accuracy of interest} (usually of larger diameter than the one of the bundle of initial conditions). When this happens, the system is unpredictable in the sense that the prediction based on any bundle of initial conditions is so imprecise that it is impossible to determine the outcome of the system with the desired prediction accuracy.\footnote{Schurz ([1996], pp.\ 133--9) discusses several variants of this form of unpredictability.} A well-known example is a system in which, due to exponential divergence of solutions, any bundle of initial conditions of at least a specific diameter spreads out over short time periods more than a diameter of interest.

The second concept of unpredictability is probabilistic.
It says that \textit{for practical purposes any bundle of initial conditions is irrelevant, i.e.\ makes it neither more nor less likely that the state is in a region of phase space of interest}. According to this concept,  it is not only impossible to predict with certainty in which region the system will be, but in addition, for practical purposes knowledge of the possible initial conditions neither heightens, nor lowers, the probability that the state is in a given region of phase space. An example is that knowledge of any bundle of sufficiently past initial conditions is practically irrelevant for predicting that the state of the system is in a region of phase space. Eagle ([2005], p.~775) defines randomness as a strong form of unpredictability: an event is random if and only if the probability of the event conditional on evidence equals the prior probability of the event. This idea relativised to practical purposes is at the heart of our second concept. Consequently, this second concept can also be regarded as a form of randomness.

Clearly, the first and second concepts of unpredictability are different and cannot be expressed in terms of each other since the notions of `diameter' and `probability' are not expressible in terms of each other.

\section{Chaos}\label{chaos}
\subsection{Defining Chaos}

I base the discussion of defining chaos on the following assumption, which is widely accepted in the literature (e.g.\ Brin and Stuck [2002], p.~23; Devaney [1986], p.~51).
A \textit{formal definition of chaos is adequate} if and only if
\begin{description}
\item (i) it captures the main \textit{pretheoretic intuitions} about chaos, and
\item (ii) it is \textit{extensionally correct} (i.e.\ correctly classifies essentially all systems which, according to the pretheoretic understanding, are uncontroversially chaotic or nonchaotic).
\end{description}

Let us first direct our attention to (i). Roughly, chaotic systems are deterministic systems showing \textit{irregular}, or even \textit{random}, behaviour and \textit{sensitive dependence to initial conditions (SDIC)}.
SDIC means that small errors in initial conditions lead to totally different solutions.

\begin{figure}
\centering
\includegraphics{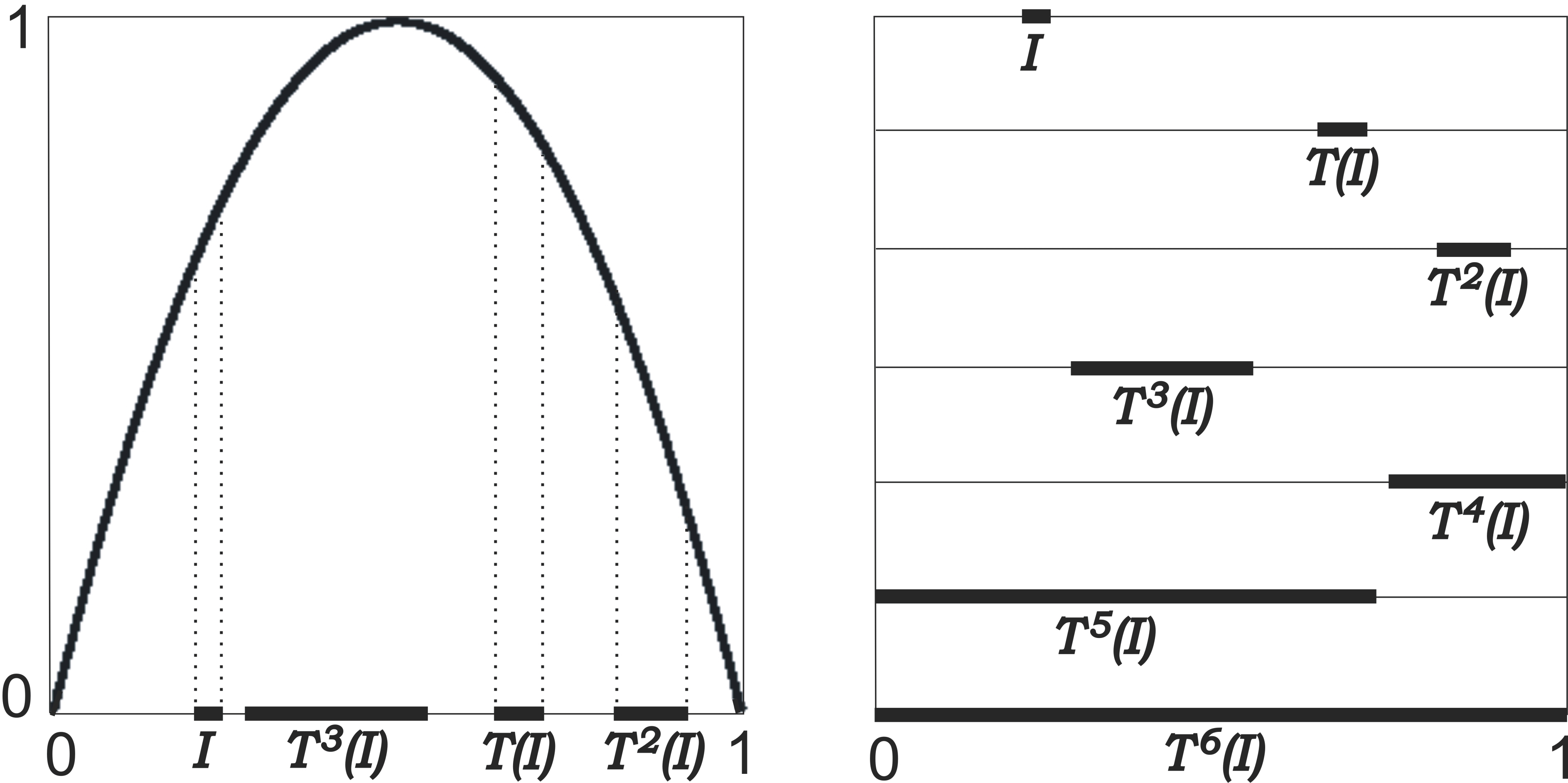}
\caption{\small{behaviour of the logistic map for $\alpha=4$}\label{LE}}
\end{figure}
The logistic map $T:[0,1]\rightarrow [0,1],\, T(x)=\alpha x(1-x)$ for $\alpha=4$ is a paradigmatic chaotic system. Figure \ref{LE} shows the first six iterates of a small bundle of initial conditions $I$, and suggests that any bundle blows up substantially. Thus the system appears to exhibit SDIC. This figure also suggests that any bundle blows up until it covers the whole phase space. Thus the motion appears not only to exhibit SDIC but also irregular behaviour in the following sense: any bundle of initial conditions eventually intersects with any other region in phase space, a property called denseness. It is widely agreed that \textit{SDIC} and \textit{denseness} are necessary conditions for chaos (Niellsen [1999], pp.~14--5; Peitgen et al.\ [1992], pp.~509--21; Smith [1998], pp.~167--9). This motivates the following criterion: a \textit{definition applying to dynamical systems captures the main pretheoretic intuitions about chaos} if and only if it implies SDIC and denseness.

Let us now discuss (ii), the requirement of extensional correctness. Imagine we are concerned with a pretheoretic property P. Further, assume that we are faced with a class of objects some of which uncontroversially have property P, others uncontroversially fail to have property P, and yet others are borderline cases or controversial in some sense. The task is to find an unambiguous definition of P. Then it is natural to say that an unambiguous definition of the property P is extensionally correct if and only if it classifies all objects correctly which uncontroversially have or do not have property P. For the borderline objects it is unimportant how they are classified, and we defer to the definition.

Being chaotic is such a property because the pretheoretic idea of chaos is somewhat vague. Among the dynamical systems whose behaviour is mathematically well understood, there is a broad class of uncontroversially chaotic systems and a broad class of uncontroversially nonchaotic systems. Moreover, there are a few borderline cases, for example the system discussed by Martinelli et al. ([1998], p.~199), where it is not clear whether they are chaotic (Brin and Stuck [2002], p.~23; Robinson [1995], pp.~81--5; Zaslavsky [2005], pp.~53--4).
Consequently, I say that a \textit{formal definition of chaos is extensionally correct} if and only if it correctly classifies essentially all mathematically well understood uncontroversially chaotic and nonchaotic behaviour.

Several definitions of chaos have been proposed (Lichtenberg and Lieberman [1992], pp.~302--9; Robinson [1995], pp.~81--6). While these definitions are very similar, they are all inequivalent. For want of space I cannot discuss all these definitions here and instead focus on a definition of chaos in terms of mixing, which will be crucial later on.

\subsection{Defining Chaos via Mixing}\label{defmix}
Intuitively speaking, the fact that a system is mixing means that any bundle of solutions spreads out in phase space like a drop of ink in a glass of water.
A measure-preserving dynamical system $(X,\Sigma,\mu, T)$ is \textit{mixing} if and only if for all $A,B \in \Sigma$:
\begin{equation}\label{SF}
\lim_{n\to \infty}\mu(T^{-n}(B)\cap A)=\mu(B)\mu(A).
\end{equation}

Mixing is occasionally mentioned in connection with chaos, usually only in the context of volume-preserving systems (e.g.\ Lichtenberg and Liebermann [1992], pp.~302--3; Schuster and Just [2005], p.~177). Yet, to the best of my knowledge, so far no one has explicitly argued that chaos can thus be defined. I will argue for this and propose that \textit{mixing is chaos}: a system is chaotic if and only if it is mixing on the relevant subset of $X$. More needs to be said about what qualifies as the relevant subset later on.

Since mixing was introduced before the 1960s, the beginning of the systematic investigation of chaos, it might seem puzzling that chaos can be adequately defined via mixing. However, many formal definitions and measures of chaos were invented before the 1960s (Dahan-Dalmedico [2004], p.~70), but rather few systems were known to which these notions apply. Novel from the 1960s onwards was that \textit{many different highly interesting systems, surprisingly also very simple systems, were found to which these concepts apply}.

Let us first discuss whether mixing captures the pretheoretic intuitions.
Mixing implies denseness: mixing systems are ergodic (Cornfeld et al.\ [1982], p.~25).
By looking at equation (\ref{ergodic}) we see that from this follows that any region, naturally interpreted as a set of positive measure, eventually visits every region in phase space.

Mixing also implies SDIC. This can be seen as follows. Mixing implies that any bundle of initial conditions spreads out uniformly over the phase space. Therefore, any bundle eventually spreads out considerably, thus exhibiting SDIC.
Formally, assume a mixing measure-preserving system $(X,\Sigma,\mu,T)$ is given where a metric $d$ is defined on $X$ and $\Sigma$ contains every open set of $X$. Further, assume that every open set has positive measure.\footnote{This is standardly assumed and, to the best of my knowledge, applies to all paradigmatic chaos systems.} Consider two open sets $O_{1}$ and $O_{2}$ with $0<\varepsilon:=\inf_{x \in O_{1}, y\in O_{2}}\{d(x,y)\}$. Mixing implies that for any open set $O$ there is a $n\geq 0$ such that $T^{n}(O)\cap O_{1}\neq \emptyset$ and $T^{n}(O)\cap O_{2}\neq \emptyset$.
But this means that $\varepsilon \leq \sup_{x,y \in T^{n}(O)}\{d(x,y)\}$. Hence the following condition holds, which in definitions like \textit{Devaney chaos} is taken to be the SDIC implied by chaotic behaviour (Devaney [1986], p.~51):
\begin{eqnarray}\label{SDIC}
\textnormal{There is a}\,\,\varepsilon>0\,\,\textnormal{such that for all}\,\, x \in X\,\,\,\textnormal{and for all}\,\,
\delta >0\,\,\,\,\,\,\,\,\,\,\,\,\,\,\,\,\,\,\,\,\,\,\,\,\,\,\,\,\,\,\,\,\,\,\,\,\,\,\\ \nonumber
\textnormal{there is a}\,\, y \in X\,\,\textnormal{and a}\,\, n \in
\field{N}_{0}\,\,\textnormal{with}\,\,d(x,y)<\delta\,\,\textnormal{and}\,\,d(T^{n}(x),T^{n}(y))\geq \varepsilon.
\end{eqnarray}

As SDIC is often linked to positive Liapunov exponents, let us now turn to a discussion of this issue. For a continuously differentiable $T$ on an open $X\subseteq \field{R}$ the \textit{Liapunov-exponent} of $x \in X$ is \begin{equation}
\lambda(x):= \lim_{n \rightarrow \infty}\frac{1}{n}\sum_{i=0}^{n-1}\log(|T'(T^{i}(x))|),
\end{equation}
where $T'$ is the derivative of $T$ (for a general definition see Ma\~{n}\'{e} [1983], p.~263). For ergodic systems the Liapunov-exponent exists and is equal for all points except for a set of measure zero (Robinson [1995], p.~86). Hence one can speak of the Liapunov-exponent of a system. Accordingly, one definition of chaos that has been suggested is that the system is ergodic and has a positive Liapunov-exponent.

From a positive Liapunov exponent it is commonly concluded that the SDIC shown by chaos consists of the exponential spreading of inaccuracies over finite time periods (e.g.\ Lighthill [1986], p.~46; Ott [2002], p.~140; Smith [1998], p.~15).\footnote{With the qualification that the time periods have to be small enough such that the inaccuracy does not eventually saturate at the diameter of the system.} However, this is mistaken. Positive Liapunov exponents imply that for almost all points $x$ in phase space the average over all $i\geq 0$ of $\log(|T'(T^{i}(x))|)$---the exponential growth rate of an inaccuracy at the point $T^{i}(x)$---is positive. Here the average is taken for the solution starting from $x$ over an \textit{infinite} time period. But positive \textit{on average} exponential growth rates over an \textit{infinite time period} do \textit{not} imply that nearby solutions diverge exponentially or rapidly over \textit{finite time periods}. The growth rate over finite time periods can be anything; inaccuracies can even shrink (Smith et al.\ [1999], pp.~2861--2).\footnote{Moreover, Liapunov exponents only measure the average growth rate of an \textit{infinitesimal} inaccuracy around $x$, which is defined as the growth rate of a small ball of radius $\varepsilon>0$ with centre $x$ as $\varepsilon\rightarrow 0$; yet in practice the uncertainty is finite and may not behave like the infinitesimal one (cf.~Bishop [unpublished], p.~8).}
Furthermore, it is not true that inaccuracies of chaotic systems spread exponentially or rapidly over finite time periods: for paradigmatic chaotic systems like the Lorenz attractor there are regions where inaccuracies even \textit{shrink} over finite time periods, and numerical evidence suggests such regions for many chaotic systems (Smith et al.~[1999], p.~2881; Zaslavsky [2005], p.~315; Ziehmann et al.~[2000], pp.~10--1).

Mixing systems need not have positive Liapunov exponents, and thus inaccuracies need not grow exponentially on average as time goes to infinity. Is this a problem for mixing as a definition of chaos? No. First, there is no agreement in the literature whether chaos should show this on average exponential growth. Some definitions do indeed demand it, others like Devaney chaos do not.
Second, the arguments for requiring positive Liapunov exponents are not convincing. The standard rationale is that the SDIC shown by chaotic system has to be exponential divergence of nearby solutions over finite time periods. But as shown above, this is not implied by a positive Liapunov exponent and also does not generally hold for chaotic systems.
Another possible argument is that for chaotic behaviour inaccuracies should spread out rapidly.
Yet the rate of divergence of mixing systems not having positive Liapunov exponents can be much faster for arbitrary long time periods than for systems with positive Liapunov exponents; thus it is not clear why positive Liapunov exponents should be required (Berkovitz et al.\ [2006], p.~689; Wiggins [1990], p.~615). To conclude, mixing captures the pretheoretic intuitions about chaos. It remains to show that mixing is extensionally correct.

To do this, I have to consider the main classes of uncontroversially chaotic and nonchaotic behaviour.\footnote{Obviously, I cannot discuss every single system regarded as clearly chaotic or nonchaotic. Yet our discussion covers all main examples.} I start with uncontroversially chaotic behaviour and first discuss volume-preserving systems. There are \textit{(i) Hamiltonian system which are chaotic on the whole hypersurface of constant energy}.
Three types of systems are mainly discussed here: first, chaotic billiards, which are mixing (Chernov and Markarian [2006]; Ott [2002], p.~296); second, hard sphere systems, which are either proven or conjectured to be mixing (Berkovitz et al.~[2006], pp.~679--80); third, geodesic flows of space with negative Gaussian curvature, which are mixing (Schuster and Just [2005], p.~181).

Another class are \textit{(ii) Hamiltonian systems to which the KAM-theorem applies}, e.g.\ the H\'{e}non-Heiles system or the standard map.
This class also includes simplified versions of Poincar\'{e} maps of systems to which the KAM-theorem applies.
The KAM-theorem describes what happens when integrable systems are perturbed by a nonintegrable perturbation. It says that tori with sufficiently irrational winding number survive the perturbation. Between the stable motion on surviving tori there appear to be regions of random motion.  As the perturbation increases, these regions become larger and often eventually cover nearly the entire hypersurface of constant energy.

For these systems the phase space is separated into regions, each of which has its own dynamics: in some of them the motion appears random and in others it is stable. Because of this separation into regions, random behaviour can only be found in a region. Consequently, as is widely acknowledged, \textit{proper chaotic motion can only occur on a region} (Ott [2002], pp.~267--95; Schuster and Just [2005], pp.~165--74). Thus I have to show that the mathematically well-understood random motion in a region is mixing. Yet the conjectured chaotic motion of KAM-type systems is understood only poorly (Zaslavsky [2005], p.~139). It has only been proven that there is chaotic behaviour near hyperbolic fixed points, where the motion is indeed mixing (Moser [1973], chapter 3).
Apart from this, some numerical evidence suggests that the motion conjectured to be chaotic is mixing (e.g.\ Chirikov [1979]). Thus Lichtenberg and Liebermann ([1992], p.~303) comment that we `expect that the stochastic orbits that we have encountered in previous sections are mixing over the bounded portion of phase space for which they exist'.

I should mention that numerical experiments suggest that for a few KAM-type maps there are sets on which the motion seems somewhat random, but these sets consist of $n\geq 2$ component areas, each of which is mapped successively on to another, returning to itself after $n$ iterations. There is no agreement whether such motion, which cannot be mixing, should be called `chaotic' (e.g.\ Belot and Earman [1997], p.~154, vs.\ Ott [2002], p.~300).
If it is, chaos can still be defined via mixing: one can say that a system is chaotic if and only if it is ergodic and its phase space is decomposable into $n\geq 1$ sets with disjoint interior such that the $n$-th iterate is mixing on each of these sets. I call this the \textit{`broad definition of chaos via mixing'}. Numerical experiments suggest that the behaviour mentioned above may be chaotic according to this definition (Ott [2002], p.~303).

Next in line are \textit{(iii) chaotic volume-preserving non-Hamiltonian systems}. Here the main examples discussed are discrete. First, the baker's map and volume-preserving Anosov diffeomorphisms like the cat map, which are mixing (Arnold and Avez [1968], p.~75; Lichtenberg and Liebermann [1992], p.~303). Second, paradigmatic chaotic systems are expanding piecewise maps like the tent map or the sawtooth map, which are mixing too (Bowen [1977]).

I now turn to dissipative systems and first discuss strange  attractors.
One class are \textit{(iv) strange attractors where the attracted solutions never enter the attractor}. Three main groups are treated here: first, for Smale's  Solenoid, and generalised Solenoid systems, there is a measure on which the motion is mixing (Mayer and Roepstorff [1983]). Second, for the system investigated by Lorenz ([1963]) and the Lorenz model, and generalised versions thereof, there is a physical measure on which the motion is mixing (Luzzatto et al.~[2005]). Third, for generalised H\'{e}non systems like the H\'{e}non map there exists a physical measure such that the motion on the attractor is mixing (Benedicks and Young [1993]).

Also important is the \textit{(v) visible chaotic behaviour of generalised logistic systems} like the logistic map. For these discrete systems for most parameter values the solutions enter an attractor with a physical measure on which the motion is either mixing or chaotic according to the broad definition via mixing. But for a few parameter values there is chaotic behaviour on the entire interval, e.g.\ for the logistic map with parameter 4; in these cases there is also a physical measure on which the motion is mixing (Jacobsen [1981]; Lyubich [2002]).\footnote{In all these cases the invariant measure is also the unique ergodic measure absolutely continuous with respect to the Lebesgue measure (Jacobsen [1981]; Lyubich [2002]).}

Finally, another class is \textit{(vi) repelling chaotic behaviour on Cantor sets}.
Two main kinds of discrete systems are discussed here: first, geometric horseshoe-systems like Smale's horseshoe, which are mixing (Robinson [1995], pp.~249--74). The second example is chaotic motion on Cantor sets for the logistic map with parameter greater than 4, which is also mixing (Robinson [1995], p.~33).\footnote{This follows because these systems are isomorphic to a Bernoulli-shift.}

Let us now turn to uncontroversially nonchaotic motion. I again start with volume-preserving systems. A paradigmatic class are \textit{(i) integrable Hamiltonian systems}, where there is periodic or quasi-periodic motion on tori, which is not mixing (Arnold and Avez [1968], pp.~210--214).

Another class is the \textit{(ii) motion on clearly nonchaotic regions of KAM-type systems}. Again, this class also includes simplified versions of Poincar\'{e} maps of KAM-type systems. As already discussed, for KAM-type systems the phase space is separated into regions, and on some regions the motion is stable. Thus I have to show that the stable motion is not mixing. And indeed, the behaviour in these regions, e.g.\ the motion on surviving tori or the one near specific elliptic periodic points, is not mixing (Arnold and Avez [1968], pp.~86--90; Lichtenberg and Liebermann [1992], chapter 3--5).

I now turn to dissipative systems. Important here are \textit{(iii) nonchaotic attractors}. These are attracting periodic cycles and fixed points and also quasi-periodic attractors as discussed by Ott ([2002], chapter 7), which obviously cannot be mixing. Moreover, the motion approaching such attractors, e.g.\ the behaviour around stable nodes or stable foci, clearly cannot be mixing (cf.~Robinson [1995], p.~105).\footnote{Here there often exists no invariant measure of interest.}

Finally, let us mention two further very broad classes of clearly nonchaotic behaviour. Since mixing captures SDIC, {\textit{(iv) systems not exhibiting any kind of SDIC}, e.g.\ the identity function, cannot be mixing.

Moreover, since mixing captures denseness, \textit{(v) motions showing SDIC but where, in any sense, typical solutions do not come arbitrarily near to any region in phase space} cannot be mixing. Examples are the system $x_{n+1}=c x_{n}$ for $c>1$ on $(0,\infty)$ or the motion around unstable nodes or unstable foci (cf.~Robinson [1995], p.~105).\footnotemark[\value{footnote}]

In sum, I have first demonstrated that mixing captures the pretheoretic intuitions about chaos. After that I have briefly shown that a definition of chaos in terms of mixing is extensionally correct in the sense explained above. Consequently, \textit{chaos can be adequately defined in terms of mixing}.

With this knowledge about chaos we are ready to critically discuss the answers suggested in the literature to our main question.

\section{Criticism of Answers in the Literature}
\subsection{New: Asymptotically Unpredictable?}\label{AU}
Let us first discuss an answer based on the concept of asymptotic unpredictability. Roughly, systems whose asymptotic behaviour cannot be predicted with arbitrary accuracy for all times, even if the bundle of initial conditions is made arbitrarily small, are said to be asymptotically unpredictable.
Let $(X,d,T)$ be a topological dynamical system, $\varepsilon$ be the desired prediction accuracy and $\delta$ be the diameter of the bundle of initial conditions. For $x\in X$ the solution $(T^{n}(x))_{n\geq 0}$ is \textit{asymptotically predictable} if and only if
\begin{equation}
\forall \varepsilon>0\,\,\exists \delta>0\,\,\forall y\in X\,\,\forall n\geq 0\,\,\,
(d(x,y)<\delta \rightarrow d(T^{n}(x),T^{n}(y))<\varepsilon).
\end{equation}
A dynamical system is \textit{asymptotically unpredictable} if and only if for all $x \nolinebreak[4] \in \nolinebreak[4]  X$ $(T^{n}(x))_{n\geq 0}$ is not asymptotically predictable.\footnote{Bishop ([2003], pp.~174--7) also aims to formalise asymptotic unpredictability. However, he does not
list the most obvious notion presented here.} In terms of the distinction introduced in subsection \ref{unpredictability}, this is clearly a version of the \textit{first} concept of unpredictability.

Miller ([1996], pp.~106--7) and Stone ([1989], p.~127) argue that the \textit{new implication of chaos for unpredictability is that chaotic systems are asymptotically unpredictable}.
Indeed, all chaotic systems discussed in the literature are asymptotically unpredictable, and standard definitions of chaos imply asymptotic unpredictability. For instance, (\ref{SDIC}), a condition of Devaney chaos and, under plausible assumptions, a consequence of mixing, clearly implies asymptotic unpredictability.

However, as Smith ([1998], p.~58) has pointed out, many nonchaotic systems, e.g.\ one only showing SDIC as it happens in the system $x_{n+1}=c x_{n}$, $c>1$, (class (v) of clearly nonchaotic behaviour), are asymptotically unpredictable. Hence this account is \textit{wrong}. But maybe the account can be strengthened in the following way: the \textit{new implication is that chaotic systems are asymptotically unpredictable and bounded}. I maintain that this is \textit{not correct} either:
there are unbounded chaotic systems (Smith [1998], pp.~168--9), a point which is reflected in usual definitions of chaos, which do not require boundedness.
Furthermore, for many bounded integrable systems (part of class (i) of the clearly nonchaotic behaviour) the solutions loop around tori in such a way that they are asymptotically unpredictable (Arnold and Avez [1968], pp.~210--4). Hence there are examples of nonchaotic, bounded and asymptotically unpredictable systems.

I conclude that the sole connection between asymptotic unpredictability and chaos is this: while only some nonchaotic systems are asymptotically unpredictable, every chaotic system is asymptotically unpredictable.

\subsection{New: Unpredictable Due to Rapid or Exponential Divergence of Solutions?}\label{URE}
It is widely believed and often claimed that the \textit{new implication of chaos for unpredictability is the following: due to rapid or exponential divergence of nearby solutions, bundles of initial conditions spread out a distance more than a diameter of interest over short time periods} (e.g.\ Ruelle [1997], pp.~27--8); \textit{often it is added that this is so and the systems are bounded} (e.g.\ Lighthill [1986], p.~46). In terms of the distinction introduced in subsection \ref{unpredictability}, this is a form of the \textit{first} concept of unpredictability.

As many unbounded nonchaotic systems like the system $x_{n+1}=cx_{n}$ with $c>1$ show (part of class (v) of clearly nonchaotic behaviour) rapid or exponentially divergence everywhere is ``nothing new'' (Smith [1998], p.~15). Thus the version not requiring boundedness \textit{cannot be true}.
But also the version requiring boundedness is \textit{wrong}: as mentioned above, there are unbounded chaotic systems. Furthermore, as argued in subsection \ref{defmix}, it is often \textit{not} true that nearby solutions of chaotic systems diverge rapidly or exponentially over finite time periods as is so widely believed in the philosophy, physics and mathematics communities (e.g.\ Eagle [2005], p.~767; Schurz [1996], p.~140; Smith [1998], p.~15). Hence this is not the sought-after new implication of chaos for unpredictability.

Why is it so widely believed that inaccuracies in chaotic systems spread rapidly or exponentially over finite time periods? One plausible reason is that because very simple systems like the cat map show this property, this claim is wrongly generalized to all chaotic systems. Also, the wrong belief stems at least in part from misinterpreting Liapunov exponents. As pointed out in subsection \ref{defmix}, positive on average exponential growth rates over an \textit{infinite time period} are wrongly taken to imply that inaccuracies spread exponentially over \textit{finite time periods}.

The only connection between the unpredictability of chaos and the rapid or exponential increase of inaccuracies over finite time periods seems to be this: it is more often the case for chaotic than for nonchaotic systems that bundles of initial conditions spread out more than a diameter of interest over short time periods.

\subsection{New: Macro-predictable \& Micro-unpredictable?}\label{macro}
Macro-predictable yet micro-unpredictable behaviour is a broad and interesting topic in physics. For instance, in statistical mechanics systems are often macro-predictable but micro-unpredictable. Here we concentrate only on whether there is any combination of macro-predictability and micro-unpredictability in chaotic systems that other deterministic systems do not have.

To gain an understanding of this third proposed answer, recall the Lorenz equations (\ref{LEO}) and Figure \ref{LA}.
These equations exhibit macro-predictability: the solutions are attracted by an attractor, a small region of phase space. There is also micro-unpredictability since the motion on the attractor exhibits SDIC. Peter Smith argues that this \textit{combination of macro-predictability and micro-unpredictability is a new implication of chaos for unpredictability}:
\begin{quote}
\small {\textit{This type of combination of large-scale order with small scale disorder, of macro-predictability with the micro-unpredictability due to sensitive dependence, is one paradigm of what has come to be called chaos.} [...]  So error inflation by itself is entirely old-hat. The novelty in the new-fangled chaotic cases that will concern us is, to repeat, the \textit{combination} of exponential error inflation with the tight confinement of trajectories by an attractor (Smith [1998], pp.~13--5, original emphasis).}
\end{quote}

Here macro-predictability means that the system eventually shows the behaviour corresponding to the motion on the attractor, a proper subset of phase space. Micro-unpredictability is understood as the unpredictability implied by exponential error inflation. Yet, as shown in section \ref{chaos}, solutions of chaotic systems need not diverge exponentially or rapidly over finite time periods. Therefore, micro-unpredictability has to be interpreted as a weaker notion, e.g.\ asymptotic unpredictability (cf.~subsection \ref{AU}).

As becomes clear from the Lorenz system, strange attractors imply this combination of macro-predictability and micro-unpredictability. However, this combination is \textit{no new implication of chaos for unpredictability} since there are many chaotic systems without attractors. As already pointed out, all chaotic volume-preserving dynamical systems like chaotic Hamiltonian systems or the baker's map (classes (i), (ii) and (iii) of uncontroversially chaotic behaviour) cannot have attractors. And some chaotic dissipative systems, e.g.\ repelling chaotic motion on Cantor sets or the logistic map on $[0,1]$ (class (vi) and a part of class (v) of uncontroverially chaotic behaviour), have no attractors. Hence these systems are \textit{not} macro-predictable in the above sense, viz.\ that appeals to attractors.

It could be that Smith ([1998]) only meant to say that this combination of macro-predictability and micro-unpredictability found in strange attractors \textit{is a novelty for systems with attractors}.
\textit{But this would not help}. Clearly, this claim would be no satisfying answer to our main question because it does not apply to essentially all chaotic systems. Furthermore, also nonchaotic systems can be macro-predictable and micro-unpredictable as discussed here.
For instance, in the plane let $R$ be the region enclosed by a circle of radius $r$ around the origin (boundary included). Imagine that all solutions in $R$ go in circles around the origin and that all solutions outside $R$ are attracted by the periodic motion in $R$ such that all solutions are continuous.
Such nonchaotic attractors (part of class (iii) of clearly nonchaotic behaviour) obviously imply macro-predictability and micro-unpredictability. Thus this combination of macro-predictability and micro-unpredictability is not even a novelty for systems with attractors.

Of course, there are also other concepts of macro-predictability and micro-unpredictability (e.g.\ Smith [1998], pp.~60--1). However, to the best of my knowledge, none of them provides a combination of macro-predictability and micro-unpredictability that is characteristic of chaotic behaviour.

To conclude, strange attractors are macro-predictable and micro-unpredictable in the above specified sense. However, it is not the case that a combination of macro-predictability and micro-unpredictability constitutes a new implication of chaos for unpredictability.

None of the answers examined so far have proven to be correct. There is one more answer suggested in the literature: some physicists, e.g.\ Ford ([1989]), have defined chaos by the condition that almost all solutions have positive algorithmic complexity. In other words they have argued that the unpredictability implied by positive algorithmic complexity is a new implication of chaos for unpredictability. However, Batterman and White ([1996]) and Smith ([1998], p.~160) have made it clear that chaos cannot be defined via algorithmic complexity since many systems without SDIC (part of class (iv) of clearly nonchaotic behaviour) have positive algorithmic complexity too. Consequently, this is \textit{no} new implication of chaos for unpredictability, and this is all we need to know.

In sum, the answers in the literature do not fit the bill.

\section{A General New Implication of Chaos for Unpredictability}
\subsection{Approximate Probabilistic Irrelevance}
The answer I propose starts from the well-known idea that mixing goes along with loss of information as recently discussed by Berkovitz et al.~([2006]). First of all, let us introduce the approximate probabilistic irrelevance, the notion of unpredictability which will be crucial for our claim.

Given a measure-preserving system $(X,\Sigma,\mu,T)$ it is common to associate with a set $A \in \Sigma$ a property $P_{A}$, where $P_{A}$ holds if and only if the system's \linebreak[4] state is in $A$ (\textit{Ibid.}, p.~671). For instance, for the logistic map with $\alpha=4$ interpreted as a model of population dynamics (May [1976]), the set $A=[0,1/2)$ corresponds to the property that the population is less than half of the maximum of the possible population.

Because time is discrete, I can denote time points by $t_{n},$ $n\in\field{Z}$, such that $n$ increases by one if the model is iterated once; for instance, if $t_{4}$ corresponds to the iteration stage $T$, $t_{5}$ corresponds to $T^{2}$ etc. Given this, I define the \textit{event} $A^{t_{n}}$ as the occurrence of the property $P_{A}$ at time $t_{n}$. To come back to our example, $A^{t_{n}}$ is the event that the population is less than half of the maximum possible population at time $t_{n}$ (Berkovitz et al.~[2006], p.~671).

Since the exact state of the system may not be known, I introduce $p(A^{t_{n}})$, the \textit{probability} of the event $A^{t_{n}}$. 
I also introduce conditional probabilities: $p(B^{t_{m}}\mid A^{t_{n}})$, for arbitrary $A$, $B \in \Sigma$ with $\mu(A)>0$, is the probability that $P_{B}$ obtains at time $t_{m}$ given that $P_{A}$ obtained at $t_{n}$ (\textit{Ibid.}, p.~671). By the usual definition, $p(B^{t_{m}}\mid A^{t_{n}})=p(B^{t_{m}}\& A^{t_{n}})/p(A^{t_{n}})$.

Now recall the second conception of unpredictability of subsection \ref{unpredictability}. For this conception we have to say what it means that  knowledge that the system is in a region $A$ at $t_{n}$ is practically irrelevant for predicting that it will be in $B$ at $t_{m}$. We say that this is so if the probability of the event $B^{t_{m}}$ given knowledge of the event $A^{t_{n}}$ \textit{approximately} equals the unconditionalised probability of the event $B^{t_{m}}$.
Let $\varepsilon>0$ be the level at which probabilities differing by less than $\varepsilon$ are considered as practically equivalent.
Further, assume that $p(A^{t_{n}})>0$; I will later explain why I am justified to do so.
Then formally this is captured by the following definition:\footnote{I use what is basically the difference measure in confirmation theory to define the approximate probabilistic irrelevance. I should point out that our claims are independent of the measure involved, i.e.\ they would remain the same if I used any other measure with the indisputable property that it is continuous when the unpredictability is highest, i.e.\ when $p(B^{t_{m}}\mid A^{t_{n}})=p(B^{t_{m}})$. Berkovitz et al.\ ([2006], p.~672) interpret the difference measure of events as a general measure of unpredictability. However, they do not justify this choice or address whether their results are independent of the measure.} 
\begin{eqnarray}\label{probrel2}
A^{t_{n}}\,\,\textnormal{is \textit{approximately probabilistically irrelevant} for predicting}\,\,B^{t_{m}}\,\, \\ \nonumber
(t_{m}\geq t_{n})\,\,\textnormal{\textit{at level}}\,\,\varepsilon> 0\,\,\textnormal{if and only if}\,\,\,
|p(B^{t_{m}}\mid A^{t_{n}})-p(B^{t_{m}})|<\varepsilon.
\end{eqnarray}

How can we determine the values of the probabilities occurring in (\ref{probrel2})?
Because the probabilities should reflect objective dynamical properties of systems, I say that the probability of an event $A^{t_{n}}$ corresponds to the measure of $A$ (\textit{Ibid.}, p.~\nolinebreak 673). As mentioned in subsection \ref{DS}, this is quite natural under certain interpretations.
\begin{equation}\label{prob1}
\textnormal{For all $t_{n}$ and for all}\,\,A\in\Sigma:\,\,p(A^{t_{n}})=\mu(A).
\end{equation}
This idea can be generalised to joint simultaneous events as follows:
\begin{equation}\label{prob2}
\textnormal{For all $t_{n}$ and for all}\,\,A,B\in\Sigma:\,\,p(A^{t_{n}}\&B^{t_{n}})=\mu(A\cap B).
\end{equation}
This implies:
\begin{equation}\label{prob3}
\textnormal{For all $t_{m},t_{n}$, $t_{m}\geq t_{n}$, and all}\,\,A,B\in\Sigma:\,\, p(B^{t_{m}}\&A^{t_{n}})=\mu(T^{n-m}(B)\cap A)
\end{equation}
since $T^{n-m}(B)$ is the evolution of the set $B$ backward in time from $t_{m}$ to $t_{n}$.\footnote{I can infer (\ref{prob3}) from (\ref{prob2}) as follows: $T^{n-m}(B)$ contains exactly those points that are in $B$ at time $t_{m}$. Consequently, $T^{n-m}(B)\cap A$ consists of exactly those points which pass $A$ at time $t_{n}$ and go through $B$ at time $t_{m}\geq t_{n}$, i.e.\ for which $B^{t_{m}}\&A^{t_{n}}$ is true. Thus from (\ref{prob2}) it follows that $p(B^{t_{m}}\&A^{t_{n}})=\mu(T^{n-m}(B)\cap A)$.}

In the next section we will see how the approximate probabilistic irrelevance relates to chaos and will finally propose an answer to our question.

\subsection{New: Sufficiently Past Events Approximately Probabilistically Irrelevant for Predictions}

The argument I put forward to answer the main question of the paper is as follows. (P1) \textit{Chaos can be defined in terms of mixing.} (P2) \textit{Mixing systems exhibit a particular pattern of approximate probabilistic irrelevance, which constitutes a form of unpredictability.}
Therefore: (C) \textit{a new implication of chaos for unpredictability is the particular pattern of approximate probabilistic irrelevance arising from mixing}.

In subsection \ref{defmix} we have seen that premise (P1) is true. Let us now argue for premise (P2). Recall the definition of mixing (\ref{SF}). I assume without loss of generality that the event we want to predict occurs at $t_{0}$. Then, assuming (\ref{prob1}) and (\ref{prob3}), it follows that a system $(X,\Sigma,\mu, T)$ is mixing if and only if
\begin{equation}\label{equivalence}
\lim_{n\rightarrow \infty} p(B^{t_{0}}\mid A^{t_{-n}})-p(B^{t_{0}})=0,
\end{equation}
for all $A,B\in\Sigma$ with $\mu(A)>0$. This equation holds for \textit{all}, i.e.\ invertible and noninvertible, measure-preserving systems. Berkovitz et al.\ ([2006], p.~676) show (\ref{equivalence})  \textit{only for invertible systems}. Moreover, they interpret their results as applying only to Hamiltonian systems. Many chaotic systems, e.g.\ all strange attractors (classes (iv) and (v) of uncontroversially chaotic behaviour), are not Hamiltonian.
Furthermore, many paradigmatic systems like generalised logistic systems or the tent map (class (v) and part of classes (iii) and (vi) of uncontroversially chaotic behaviour) are not invertible. Since I am interested in the unpredictability implied by chaos, I
need (\ref{equivalence}) for all systems, and this general claim follows from (\ref{prob1}) and (\ref{prob3}).

From the definition of the limit, I obtain that (\ref{equivalence}) can be expressed as:
\begin{eqnarray}
\textnormal{For any event}\,\,B^{t_{0}},\,\,\textnormal{any precision}\,\,\varepsilon>0
\,\,\textnormal{and any}\,\,A\,\,\textnormal{with}\,\,\mu(A)>0\,\,\,\,\,\,\,\,\,\,\,\,\,\,\,\,\,\,\,\,\,\,\,\, \\ \nonumber
\textnormal{there exists}\,\,n_{0}\in\field{N}\,\,\textnormal{such that for all}\,\,n\geq n_{0}:|p(B^{t_{0}}\mid A^{t_{-n}})-p(B^{t_{0}})|< \varepsilon.\,\,\,
\end{eqnarray}
Hence mixing means that \textit{for predicting an arbitrary event at an arbitrary level of precision $\varepsilon>0$, any sufficiently past event is approximately probabilistically irrelevant}. Notice that due to the impossibility of determining initial conditions precisely, scientists always consider regions of phase space corresponding to possible initial conditions. Since these regions are not of measure zero, I am justified assuming that $\mu(A)>0$. In terms of the distinction introduced in subsection \ref{unpredictability}, this pattern of probabilistic irrelevance is a version of the \textit{second} concept of unpredictability. Hence mixing systems exhibit a particular pattern of approximate probabilistic irrelevance, which constitutes a form of unpredictability: i.e.\ premise (P2) is true.\footnote{
This claim can be generalised. $(X,\Sigma,\mu,T)$ is mixing iff for any $\rho$ absolutely continuous with respect to $\mu$ and any square integrable function $f$: $\lim_{n\rightarrow \infty}\int f(x)d\rho_{n}=\int f(x) d\mu$,
where $\rho_{n}$ is the $n$-steps evolved measure.
Interpret $\mu$ as probability and $\rho$ as measuring our knowledge of the initial condition. Then, assuming absolute continuity of $\rho$, mixing means that for arbitrary knowledge of the initial condition after a sufficiently long time the prediction obtained by evolving the measure is practically no better than if we had no knowledge whatsoever of the initial conditions (cf.~Berger [2001], pp.~126--32).}

Now that I have argued for the premises (P1) and (P2) of the above argument, I conclude: (C) \textit{ a general new implication of chaos for unpredictability is that for predicting any event at any level of precision $\varepsilon>0$, all sufficiently past events are approximately probabilistically irrelevant.}

To fully understand this conclusion, consider the following: for strange attractors this claim applies in a strict sense only to events on the attractor. Yet for practical matters there is chaotic behaviour when solutions are very near to the strange attractor (cf.~subsection \ref{DS}); then my claim means that for predicting any event \textit{on or very near the attractor $\Lambda$} at any level of precision $\varepsilon>0$, all sufficiently past events \textit{in the neighbourhood $U\supset \Lambda$} are approximately probabilistically irrelevant. For KAM-type systems my claim applies, as one would like it, to each chaotic region.
Moreover, as explained in section \ref{defmix} in discussing the uncontroversially chaotic behaviour, some may want to adopt the broad definition of chaos via mixing, i.e.\ that the system is ergodic and its phase space is decomposable into $n\geq 1$ regions with disjoint interior such that the $n$-th iterate is mixing on each set. When $n>1$, my claim (C) has to be adapted in the following way: the unpredictability of mixing applies to the $n$-th iterate on the region of interest. This means that for predicting any event \textit{in the region of interest} at any level of precision $\varepsilon>0$, all sufficiently past events that \textit{could have evolved to the region of interest} are approximately probabilistically irrelevant.

On the one hand, the unpredictability involved in my answer is strong: sufficiently distant events are \textit{practically as independent} as coin tosses. On the other hand, it is weak since only \textit{sufficiently} past measurements are approximately probabilistically irrelevant. Restricting my claim to sufficiently past events is essential: first, many chaotic systems are continuous, and continuity makes it impossible that for all past times, all events are approximately probabilistically irrelevant for predictions. Second, we have seen that to require rapid divergence of nearby solutions for chaotic behaviour is untenable.

What is novel about my claim? Granted, in a few publications on chaos the notion of `irrelevance' is discussed. In fact, there are two main foci; but none give my claim. First, there is Berkovitz et al.'s ([2006]) explication of the ergodic hierarchy. Yet recall our main argument (cf.~the beginning of this subsection). As pointed out, Berkovitz et al.\ only show premise (P2) for invertible systems, and they interpret their results as only applying to Hamiltonian systems.
Hence they do not argue for the general premise (P2), and, most importantly, they do not argue for the crucial premise (P1). Therefore, they could not arrive at the conclusion (C).
Second, sometimes it is asserted that for chaos the input is irrelevant in the sense that prediction is
exponentially expensive in the initial data, meaning that for an input string of length $n$ all information is lost after $n$ steps, at which point we are totally unsure what happens next (Leiber [1998], p.~361; Smith [1998], p.~53). However, as argued in subsection \ref{URE}, predictions for chaotic systems need not be exponentially expensive in the initial data; the irrelevance shown by chaos is more subtle.

\section{Conclusion}
The unpredictability of chaotic systems is one of the issues that has attracted most interest in chaos research.
Nonetheless, nearly half a century after the start of the systematic investigation of chaos, there has been much confusion about, and no correct answer to, the question `What are the new implications of chaos for unpredictability?', in the sense that chaotic systems are unpredictable in a way that other deterministic systems are not.

I have criticised the answers in the literature to the above question. First, I rejected the answer that chaotic systems are asymptotically unpredictable on the grounds that also many nonchaotic systems are asymptotically unpredictable. Second, I rejected the answer that chaotic systems are unpredictable in the sense of exponential or rapid divergence of nearby solutions (often claimed with the added  condition of boundedness). For, when not requiring boundedness, many nonchaotic systems are also unpredictable in this sense. Furthermore, in the case of requiring boundedness, there are unbounded chaotic systems and, though unacknowledged in the philosophy literature, chaotic systems need not be unpredictable in the sense of having exponential or rapid divergence of solutions. Third, I dismissed the answer that chaos shows a specific combination of macro-predictability and micro-unpredictability: there are chaotic systems which are not macro-predictable and nonchaotic systems which also show this combination of macro-predictability and micro-unpredictability.

This prompted the search for an alternative answer. I approached this problem by showing that chaos can be defined in terms of mixing, i.e.\ that mixing captures the main pretheoretic intuitions about chaos and correctly classifies the various classes of uncontroversially chaotic and nonchaotic behaviour. This has never been explicitly argued for in the literature. Based on this insight, I proposed a novel general answer: a new implication of chaos for unpredictability is that for predicting any event at any level of precision $\varepsilon>0$, all sufficiently past events are approximately probabilistically irrelevant. Chaotic behaviour is multi-faceted and takes various forms. Yet if the aim is to identify a general new implication of chaos for unpredictability, I think this is the best we can get.

\section*{Acknowledgments}
I am indebted to Jeremy Butterfield, Roman Frigg, Peter Smith and two anonymous referees for valuable comments on earlier versions of this paper. Many thanks also to Robert Bishop, Adam Caulton, Franz Huber, Paul Weingartner and the audiences at the Philosophy Workshop at the University of Cambridge and the 5th UK and European meeting on the Foundations of Physics for discussions that lead to improvements in this paper. I am grateful to St John's College, Cambridge, for financial support.

\section*{References}
\addcontentsline{toc}{section}{References}
\begin{list}{}{%
    \setlength{\labelwidth}{0pt}
    \setlength{\labelsep}{0pt}
    \setlength{\leftmargin}{24pt}
    \setlength{\itemindent}{-24pt}
  }
\item Arnold, V. I. and Avez, A. [1968]: \textit{Ergodic Problems of Classical Mechanics}, New York: W.A. Benjamin.

\item Batterman, R. W. and White, H. [1996]: `Chaos and Algorithmic Complexity', \textit{Foundations of Physics}, \textbf{26}, pp.~307--37.

\item Belot, G. and Earman, J. [1997]: `Chaos Out of Order: Quantum Mechanics, the Correspondence Principle and Chaos', \textit{Studies in History and Philosophy of Modern Physics}, \textbf{28}, pp.~147--82.

\item Benedicks, M. and Young, L.-S. [1993]: `Sinai-Ruelle-Bowen Measures for Certain H\'{e}non Maps', \textit{Inventiones Mathematicae}, \textbf{112}, pp.~541--76.

\item Berger, A. [2001]: \textit{Chaos and Chance, An Introduction to Stochastic Aspects of Dynamics}, New York: de Gruyter.

\item Berkovitz, J., Frigg, R. and Kronz, F. [2006]: `The Ergodic Hierarchy, Randomness and Hamiltonian Chaos', \textit{Studies in History and Philosophy of Modern Physics}, \textbf{37}, pp.~661--91.

\item Bishop, R. C. [2003]: `On Separating Predictability and Determinism', \textit{Erkenntnis}, \textbf{58}, pp.~169--88.

\item Bishop, R. C. [unpublished]: `What Could Be Worse Than the Butterfly Effect?'.

\item Bowen, R. [1977]: `Bernoulli Maps of the Interval', \textit{Israel Journal of Mathematics}, \textbf{28}, pp.~161--8.

\item Brin, M. and Stuck, G. [2002]: \textit{Introduction to Dynamical Systems}, Cambridge: Cambridge University Press.

\item Chernov, N. and Markarian, R. [2006]: \textit{Chaotic Billiards}, Providence: American Mathematical Society.

\item Chirikov, B. V. [1979]: `A Universal Instability of Many-Dimensional Oscillator Systems', \textit{Physics Reports}, \textbf{52}, pp.~264--379.

\item Cornfeld, I. P., Fomin, S. V. and Sinai, Ya. G. [1982]: \textit{Ergodic Theory}, Berlin et al.: Springer.

\item Dahan-Dalmedico, A. [2004]: `Chaos, Disorder, and Mixing: a New Fin-de-si\`{e}cle Image of Science?', in M. N. Wise (ed), 2004, \textit{Growing Explanations, Historical Perspective on the Sciences of Complexity}, Durham: Duke University Press, pp.~67--94.

\item Eagle, A. [2005]: `Randomness is Unpredictability', \textit{The British Journal for the Philosophy of Science}, \textbf{56}, pp.~749--90.

\item Earman, J. [1971]: `Laplacian Determinism, or Is This Any Way to Run a Universe?', \textit{Journal of Philosophy}, \textbf{68}, pp.~729--44.

\item Eckmann, J.-P. and Ruelle, D. [1985]: `Ergodic Theory of Chaos and Strange Attractors', \textit{Reviews of Modern Physics}, \textbf{57}, pp.~617--54.

\item Ford, J. [1989]: `What is Chaos That We Should Be Mindful of It?', in P. Davies (ed), 1989, \textit{The New Physics}, Cambridge: Cambridge University Press, pp.~348--71.

\item Jacobsen, M. V. [1981]: `Absolutely Continuous Invariant Measures for One-Parameter Families of One-Dimensional Maps', \textit{Communications in Mathematical Physics}, \textbf{81}, pp.~39--88.

\item Leiber, T. [1998]: `On the Actual Impact of Deterministic Chaos', \textit{Synthese}, \textbf{113}, pp.~357--79.

\item Lichtenberg, A. J. and Lieberman, M. A. [1992]: \textit{Regular and Chaotic Dynamics}, Berlin et al.: Springer.

\item Lighthill, J. [1986]: `The Recently Recognized Failure of Predictability in Newtonian Dynamics', \textit{Proceedings of the Royal Society of London, Series A}, \textbf{407}, pp.~35--50.

\item Lorenz, E. N. [1963]: `Deterministic Nonperiodic Flow', \textit{Journal of the Atmospheric Sciences}, \textbf{20},  pp.~130--41

\item Luzzatto, S., Melbourne, I. and Paccaut, F. [2005]: `The Lorenz Attractor is Mixing', \textit{Communications in Mathematical Physics}, \textbf{260}, pp.~393--401.

\item Lyubich, M. [2002]: `Almost Every Quadratic Map is Either Regular or Stochastic', \textit{Annals of Mathematics}, \textbf{156}, pp.~1--78.

\item Malament, D. B. and Zabell, S. L. [1980]. `Why Gibbs Phase Averages Work---The Role of Ergodic Theory', \textit{Philosophy of Science}, \textbf{47}, pp.~339--49.

\item Ma\~{n}\'{e}, R. [1983]: \textit{Ergodic Theory and Differentiable Dynamics}, Berlin et al.: Springer.

\item Martinelli, M., Dang, M. and Seph, T. [1998]:
`Defining Chaos', \textit{Mathematics Magazine}, \textbf{71}, pp.~112--22.

\item May, R. M. [1976]: `Simple Mathematical Models with Very Complicated
Dynamics', \textit{Nature}, \textbf{261}, pp.~459--67.

\item Mayer, D. and Roepstorff, G. [1983]: `Strange Attractors and Asymptotic Measures of Discrete-Time Dissipative Systems', \textit{Journal of Statistical Physics}, \textbf{31}, pp.~309--26.

\item Miller, D. [1996]: `The Status of Determinism in an Uncontrollable World', in P. Weingarnter and G. Schurz (eds), 1996, \textit{Law and Prediction in the Light of Chaos Research}, Berlin et al.: Springer, pp.~103--14.

\item Montague, R. [1962]: `Deterministic Theories', in D. Wilner (ed), 1962, \textit{Decisions, Values and Groups}, New York: Pergamon Press, pp.~325--70.

\item Moser, J. [1973]: \textit{Stable and Random Motions in Dynamical Systems}, Princeton: Princeton University Press.

\item Nillsen, R. [1999]: `Chaos and One-to-Oneness', \textit{Mathematics Magazine}, \textbf{72}, pp.~14--21.

\item Ott, E. [2002]: \textit{Chaos in Dynamical Systems}, Cambridge: Cambridge University Press.

\item Peitgen, H.-O., J\"{u}rgens, H. and Saupe, D. [1992]: \textit{Chaos and Fractals, New Frontiers of Science}, New York: Springer.

\item Robinson, C. [1995]: \textit{Dynamical Systems: Stability, Symbol Dynamics, and Chaos}, London: CRC Press.

\item Ruelle, D. [1997]: `Chaos, Predictability, and Idealizations in Physics', \textit{Complexity}, \textbf{3}, pp.~26--8.

\item Schurz, G. [1996]: `Kinds of Unpredictability in Deterministic Systems', in P. Weingarnter and G. Schurz (eds), 1996, \textit{Law and Prediction in the Light of Chaos Research}, Berlin et al.: Springer, pp.~123--41.

\item Schuster, G. and Just, W. [2005]: \textit{Deterministic Chaos, An Introduction}, Weinheim: Wiley-VCH Verlag.

\item Smith, L. A., Ziehmann, C. and Fraedrich, K. [1999]: `Uncertainty Dynamics and Predictability in Chaotic Systems', \textit{Quarterly Journal of the Royal Meteorological Society}, \textbf{125}, pp.~2855--86.

\item Smith, P. [1998]: \textit{Explaining Chaos}, Cambridge: Cambridge University Press.

\item Stone, M. A. [1989]: `Chaos, Prediction and Laplacian Determinism', \textit{American Philosophical Quarterly}, \textbf{26}, pp.~123--31.

\item van Lith, J. [2001]: `Ergodic Theory, Interpretations of Probability, and the Foundations of Statistical Mechanics', \textit{Studies in History and Philosophy of Modern Physics}, \textbf{32}, pp.~581--95.

\item Weingartner, P. [1996]: `Under What Transformations Are Laws Invariant?',
in P. Weingarnter and G. Schurz (eds), 1996, \textit{Law and Prediction in the Light of Chaos Research}, Berlin et al: Springer, pp.~47--88.

\item Wiggins, S. [1990]: \textit{Introduction to Applied Nonlinear Dynamical Systems and Chaos}, Berlin et al.: Springer.

\item Young, L.-S. [1997]: `Ergodic Theory and Chaotic Dynamical Systems,' \textit{XII-th International Congress of Mathematical Physics (Brisbane)}, Cambridge, MA: International Press, pp.~311--9.

\item Young, L.-S. [2002]: `What Are SRB Measures, and Which Dynamical Systems Have Them?', \textit{Journal of Statistical Physics}, \textbf{108}, pp.~733--54.

\item Zaslavsky, G. M. [2005]: \textit{Hamiltonian Chaos and Fractional Dynamics}, Oxford: Oxford University Press.

\item Ziehmann, C., Smith, L. A. and Kurths, J. [2000]: `Localized Lyapunov  Exponents and the Prediction of Predictability', \textit{Physics Letters A}, \textbf{271}, pp.~1--15.

\end{list}

\end{document}